\documentclass[apsprb,english]{revtex4-1}

\usepackage{amsmath}
\usepackage{textcomp}
\usepackage[pdftex]{graphicx}
\usepackage{xr} 
\usepackage{float}
\usepackage[lmargin=50pt,rmargin=50pt,tmargin=50pt,bmargin=50pt]{geometry}

\usepackage[usenames]{color}

\begin{document}

	\title{Nanosecond Protein Dynamics in a Red/Green Cyanobacteriochrome Revealed by Transient IR Spectroscopy}
\author{David Buhrke*, Kerstin T. Oppelt, Philipp J. Heckmeier, Ricardo Fernández-Terán and Peter Hamm \\\textit{Department of Chemistry, University of
		Zürich, Zürich, Switzerland}\\  $^*$david.buhrke@chem.uzh.ch
}
\date{\today}

\begin{abstract}
	\noindent Over the last decades, photoreceptive proteins were extensively studied with biophysical methods to gain a fundamental understanding of their working mechanisms and further guide the development of optogenetic tools. Time-resolved infrared (IR) spectroscopy is one of the key methods to access their functional non-equilibrium processes with high temporal resolution, but has the major drawback that experimental data is usually highly complex. Linking the spectral response to specific molecular events is a major obstacle. Here, we investigate a cyanobacteriochrome (CBCR) photoreceptor with a combined approach of transient absorption spectroscopy in the Visible and IR spectral regions. We obtain kinetic information in both spectral regions by analysis with two different fitting methods: global multiexponential fitting and lifetime analysis. We investigate the ground state dynamics that follow photoexcitation in both directions of the bi-stable photocycle (Pr* and Pg*) in the nanosecond and microsecond time regime. We find two ground state intermediates associated with the decay of Pr* and four with Pg* and report the macroscopic time constants of their interconversions.  One of these processes is assigned to a structural change in the protein backbone. 
\end{abstract}

\maketitle

\section{Introduction}

Cyanobacteriochromes (CBCRs) are modular light-regulated enzymes that adjust the activity of catalytic output modules (e.g. a histidine kinase domain) through bi-stable photoswitching of one or more distant photosensory modules (PSM) \cite{Rockwell2010,Rockwell2017,Fushimi2019}. Thus, CBCRs enable cyanobacteria to regulate diverse cellular processes in response to changes in environmental light conditions. The PSMs adapt a GAF (cGMP-phosphodiesterase/adenylate cyclase/FhlA) fold motif and bind various bilin chromophores, such as phycocyanobilin (PCB), to one or more cysteine residues. CBCR PSMs are becoming increasingly popular as building blocks for optogenetic tools, because they can be fused to a variety of catalytic domains to allow for allosteric photocontrol of their respective reactions. To name a few examples, CBCRs recently have been engineered to function as adenylyl cyclases \cite{Blain-Hartung2018, Hu2018}, to control protein-protein interactions \cite{Jang2019}, and gene expression in bacteria \cite{Ong2018,Castillo-Hair2019}. Furthermore, the small size, spectral diversity and fluorescence properties of CBCR PSMs are promising for applications as markers in fluorescence multiplexing and super-resolution microscopy \cite{Oliinyk2019}.
Most of the biotechnological applications considered introduce directed modifications of the respective PSMs, and thus detailed knowledge about the PSM structure and dynamics on a molecular level is required. While X-ray crystallography \cite{Narikawa2013,Burgie2013} and NMR spectroscopy \cite{Lim2018} provide structural information for thermally stable (parent) states of the CBCRs, transient spectroscopies are valuable tools to study the photoinduced non-equilibrium dynamics \cite{Slavov2015,Fukushima2011,Hardman2014,Hauck2014,Choudry2018,Lorenz-fonfria, Hardman2020}.

\begin{figure}[ht]
	\centering
	\includegraphics[width=.5\linewidth]{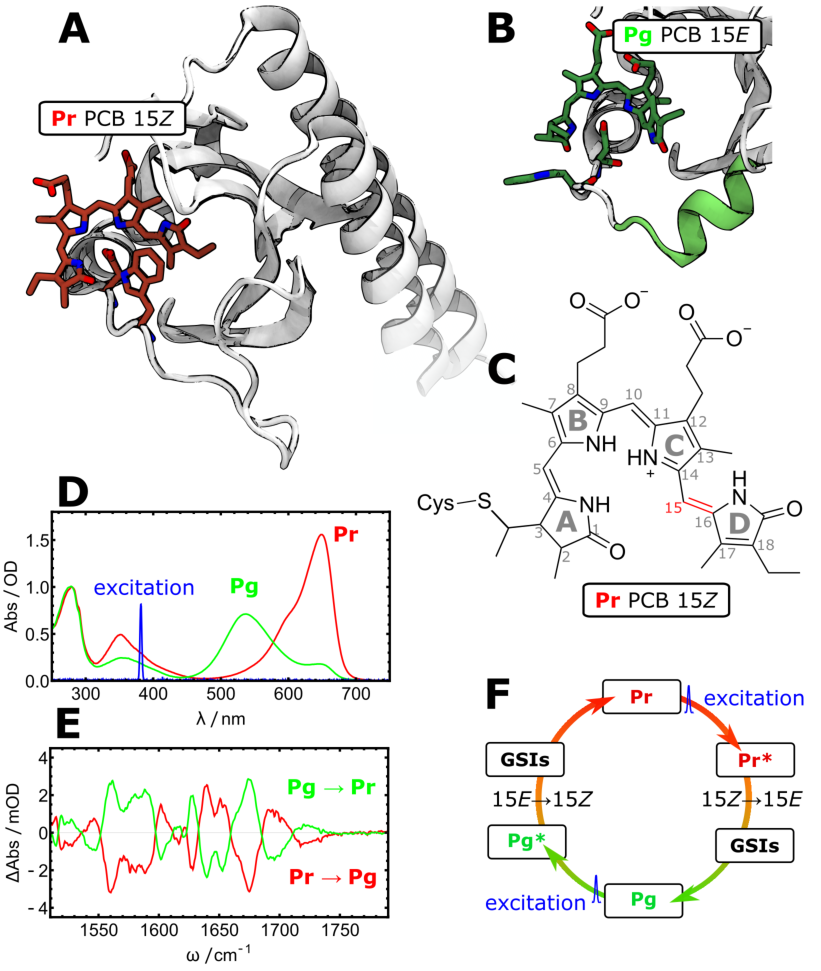}
	\caption{Properties of the photosensory GAF domain \mbox{Slr-g3}. \textbf{A} Crystal structure in the red-light absorbing Pr state (PDB entry 5DFX). \textbf{B} Detailed view of the chromophore binding pocket in the green-light absorbing Pg state (PDB entry 5M82), highlighting structural differences between Pr and Pg. \textbf{C} Nomenclature of the phycocyanobilin chromophore, depicted in the 15\textit{Z} configuration. The four pyrrole rings (A-D) are connected by methine bridges (C5, C10, C15). \textbf{D} Steady state UV-Vis spectra (Pr in red and Pg in green). Blue trace: spectrum of the 380~nm pulses used for excitation in the transient experiments. \textbf{E} FT-IR difference spectra, red: Pr $\rightarrow$ Pg transition, green: vice versa. \textbf{F} Photocycle scheme of Slr-g3: after excitation (*) of the Pr and Pg states, the chromophore isomerizes, the system decays to the electronic ground state and proceeds to form the respective other parent state via a series of GSIs. The transient experiments in this study cover the time window up to 42 \textmu s, but the reactions are finished only after 10 ms.}
	\label{intro}
\end{figure}

Here, we investigate a PSM of the CBCR Slr1393 from \textit{Synechocystis} PCC6803 by transient absorption spectroscopy in the infrared (IR) and visible spectral regions. This protein domain binds PCB and is located in a sequence after two non-photoactive GAF domains (counted from the N-terminus), therefore denoted Slr-g3 in the following. Slr1393 naturally acts as a light-regulated histidinee kinase, but by artificially  fusing Slr-g3 to an adenylyl cyclase, photocontrol over this domain was also achieved \cite{Hu2018}. Slr-g3 is a small protein domain (190~aa, 22~kDa) that converts reversibly between a red-light absorbing (Pr) and a green-light absorbing (Pg) parent state \cite{Xu2014}. The crystal structures in both parent states were recently solved (Fig. \ref{intro}A and B) \cite{Xu2020}, and display distinct structural differences of PCB and the protein backbone. In the Pr state, PCB is found in the 15\textit{Z} configuration (Fig. \ref{intro}C), closely resembling the Pr state of phytochromes. In the Pg state on the other hand, one methine bridge is isomerized (15\textit{E}), and the outer pyrrole rings A and D are twisted out of plane, leading to an effective reduction in conjugation length and thus the characteristic hypsochromic shift of the absorption maximum (Fig. \ref{intro}D) \cite{Wiebeler2018,Wiebeler2019,Buhrke2020,Xu2020}. The Pg and Pr states differ also with respect to the protein structure, e.g. the position of a tryptophan residue and $\alpha$-helicity (the latter highlighted in Fig. \ref{intro}B). The differences in the chromophore and protein configuration in Pr and Pg are reflected in changes of the Vis and IR absorption spectra (Fig. \ref{intro}D and E).
Similar to other photoreceptor proteins, the light-induced reactions of Slr-g3 may be described by a photocycle scheme (Fig. \ref{intro}F), where electronically excited states decay via a series of ground state intermediates (GSIs) to the respective product states. The ultrafast excited state dynamics have been studied in detail by transient Vis spectroscopy \cite{Slavov2015}, and we extend this approach by employing transient Vis and IR spectroscopy in parallel to study the photocycle reactions in a time window between 100~ps and 42~\textmu s. We analyze the data with two different fitting schemes to understand the complex kinetic information in the Vis and IR regions and relate it to the coupling of the transitions in the chromophore and the protein environment.

\section{Materials and Methods}

\subsection*{Protein Expression and Purification}
Slr-g3 was expressed and assembled with PCB in \textit{Escherichia coli} BL21 cells in darkness. The holo-Slr-g3 expressing cell line \cite{Buhrke2020} was a generous gift from the lab of Thomas Friedrich (TU Berlin). The protein was purified under native conditions via Ni-affinity chromatography and a His$_{6}$-Tag N-terminal to the Slr-g3 domain and desalted using a Sephadex HiPrep 26/10 column (GE Healthcare Bio-Sciences, Uppsala, Sweden) into a final buffer containing 50~mM Tris (pH~7.8 or 7.4 for subsequent D$_{2}$O exchange), 300~mM NaCl and 5~mM EDTA. For all IR experiments, the samples were prepared in D$_{2}$O buffer using the following protocol: samples were lyophilized, re-dissolved in D$_{2}$O and kept in the dark at 4°C for at least 5~h to ensure complete H/D exchange. The samples were lyophilized again for storage and dissolved in D$_{2}$O only immediately before the measurements. The integrity of the samples during the lyophilization steps was monitored by recording UV-Vis spectra of the Pr and Pg states before and after the procedure and no differences were found (Fig. S1).

\subsection*{Spectroscopy} Steady-state UV-Vis spectra were recorded with a Nanodrop 1000 Spectrophotometer (Thermo Scientific, Waltham, MA, USA) and a UV-2450 UV-Vis Spectrophotometer (Shimadzu, Nakagyo-ku, Kyoto, Japan). The samples were prepared either in the Pr or Pg states by illumination with green and red LED arrays (LIU630A and LIU525B, Thorlabs, Newton, MA, USA) as starting points for all experiments. FT-IR difference spectra were recorded with a Tensor 27 spectrometer (Bruker, Ettlingen, Germany).
 In all time-resolved Vis and IR experiments the samples were prepared under the same conditions to ensure maximum comparability of the datasets. The sample, with a concentration of 0.7~mM (computed with the ExPASy ProtParam tool from the sequence and $\mathrm{OD_{280}}$(0.1 mm)=~0.25), was cycled in a closed system with a peristaltic pump to ensure sufficient sample exchange in the probe spot. The fresh sample conditions were checked at time delays before t0, and traces recorded $-10~ns$ were used for background subtraction. This closed system included a reservoir where the sample was illuminated with the same LED arrays that were used in the static experiments to prepare either the Pr or Pg state and was constantly purged with $\mathrm{N_2}$. The measurement cell consisted of two 2-mm thick $\mathrm{CaF_2}$ windows separated by a 50~\textmu m Teflon spacer. Transient Vis and IR experiments employed a pump-probe scheme with two electronically-synchronized Ti:Sapphire laser systems running at 2.5~kHz \cite{Bredenbeck2004}. Pump-probe difference spectra were obtained by alternately blocking consecutive pump laser shots using a mechanical chopper and acquired up to the maximum delay value of 42~\textmu s with the same delay times. The time resolution was limited by the length of the pump pulse (60~ps, while the synchronization jitter of the setup was 10~ps). The linear polarization of the pump pulse was set to magic angle (54.7$^{\circ}$) relative to the p-polarized probe pulse. A multichannel referencing scheme was used to suppress noise in all transient experiments \cite{Feng2017}.

For the UV pump pulses, the respective laser was tuned to 760~nm, such that second harmonic generation in a BBO crystal produced pulses with a center wavelength of 380~nm (see Fig. \ref{intro}D). The compressor stage of the amplifier was bypassed, and stretched pulses of ca. 60~ps FWHM duration (determined on the rising absorption edge on a silicon wafer) and a power of 3~\textmu J were employed to ensure mild pumping conditions and minimize sample degradation (the spot size was ca. $140~\mu m $ FWHM). The visible probe pulses were generated by tightly focusing ca. 1~\textmu J of the 800~nm pulses generated by the probe laser into a stationary 3~mm thick sapphire plate. After passing the sample, the probe beam was collimated, dispersed by a UV transmission grating (Thorlabs, 830~mm$^{-1}$), focused by 75~mm fused silica lenses onto a  2048-pixel CMOS line array (Synertronic Designs). The probe spectral axis was calibrated by fitting the position of the transmission maxima of several interference filters, and the light intensity was controlled by using broadband neutral density filters to prevent saturation of the detector. Color-balancing filters were used to homogenize the light intensity profile of the probe and reference beams, and to filter out stray pump light. The obtained spectral resolution was ca. 0.8~nm after binning of four adjacent pixels to improve the signal-to-noise ratio, yielding 512 effective pixels.

Mid-IR probe pulses centered at 1600 or 1720~cm$^{-1}$ (duration ca. 100~fs) were generated in an optical parametric amplifier (OPA) \cite{Hamm2000}, passed through a spectrograph and detected in a 2$\times$64 MCT array detector with a spectral resolution of $\mathrm{\approx}$ 2~cm$^{-1}$/pixel. The two spectral regions had an overlap of ca. 20~cm$^{-1}$ which was used to join the transient spectra, and hence obtain a representation spanning the range from 1520 to 1780~cm$^{-1}$. An FT-IR spectrum of water vapor and the water vapor lines of the non-purged setup were used for frequency calibration.

\section{Data Analysis}

All data sets were analyzed with two complementary methods: lifetime analysis and global multiexponential fitting. The fundamental assumption behind both methods is that the system under investigation can be described by interconverting discrete states with time-invariant spectra.
That is, the data matrix $d(\omega_i,t_j)$ can be written as superposition of the $n$ different components:
\begin{equation}
d(\omega_i,t_j) = \sum_{k=1}^{n}C_k(t_j)A_k(\omega_i),
\label{separability}
\end{equation}
where $C_k(t_j)$ is the concentration profile of component $k$ as a function of time $t_j$, and $A_k(\omega_i)$ its spectrum at probe frequency $\omega_i$. For both methods, the kinetic traces were fit to multiexponential functions \cite{Hobson1998,Kumar2001,Lorenz-Fonfria:06}:
\begin{equation}
f(\omega_i,t_j)=a_0(\omega_i)-\sum_{k=1}^n a(\omega_i,\tau_k)e^{-t_j/{\tau_k}},
\label{LDA}
\end{equation}
where the index $k$ refers to a kinetic component with time constant $\tau_k$. We will abbreviate $a_{i,k}\equiv a(\omega_i,\tau_k)$.

\subsection*{Lifetime analysis}
For the lifetime analysis, the time constants $\tau_k$ were fixed and distributed equidistantly on a logarithmic scale with 10 terms per decade, while only the amplitudes $a_{i,k}$ were the free fitting parameters. A penalty function that maximizes the generalized absolute Shannon–Jaynes entropy $s_i$ of the amplitudes $a_{i,k}$ for each frequency $i$ was introduced to regularize the fit and avoid overfitting \cite{Lorenz-Fonfria:06,Lorenz-Fonfria2007}:
\begin{equation}
\begin{aligned}
s_i=\sum_{k}\Bigg(& \sqrt{a_{i,k}^2+4m_i^2}\\&-a_{i,k} \mathrm{ln}\frac{\sqrt{a_{i,k}^2+4m_i^2}+a_{i,k}}{2m_i}-2m_i\Bigg)
\end{aligned}
\label{Entropy}
\end{equation}
Here, $m_i$ are the so-called \textit{a priori} solutions, which are a measure for the overall amplitude of the data at $\omega_i$ \cite{Lorenz-Fonfria:06,Lorenz-Fonfria2007}. The entropy is subtracted from the root mean square deviation $\chi^2$ of the fit,
weighted by a regularisation parameter $\lambda$:
\begin{equation}
E_i=\frac{\chi^2_i}{\lambda} - s_i,
\label{error}
\end{equation}
with
\begin{equation}
\chi^2=\sum_{i,j}\left(d(\omega_i,t_j)-f(\omega_i,t_j)\right)^2.
\label{chisquare}
\end{equation}
The metric $E_i$ is minimized with respect to the amplitudes $a_{i,k}$. Two different criteria were tested to select the regularisation parameter $\lambda$\cite{Lorenz-Fonfria2007}. Here, the discrepancy criterion yielded reasonable lifetime fits, while the alternative approach of regularisation with the L-curve criterion resulted in clear overfitting (see SI for details).

This type of lifetime analysis yields lifetime density maps (e.g. Fig. \ref{Fig2} B) instead of discrete time constants. The procedure has been tested extensively in the literature with different synthetic and experimental data sets where the impact of parameters such as $\lambda$ is described in detail\cite{Lorenz-Fonfria:06,Lorenz-Fonfria2007,Lorenz-Fonfria2015}. It is also available as part of the software packages OPTIMUS\cite{Slavov2015a} or PyLDM\cite{Dorlhiac2017}.

The sum over the squared amplitudes at all probe frequencies at one time $\tau_k$ is termed the ``dynamical content'' $D(\tau_k)$\cite{Stock2018,Bozovic2020}:
\begin{equation}
D(\tau_k) =\sqrt{\sum_{i} a^2_{i,k}}.
\label{Dynamical content}
\end{equation}

\begin{figure}[h]
	\centering
	\includegraphics[width=.5\linewidth]{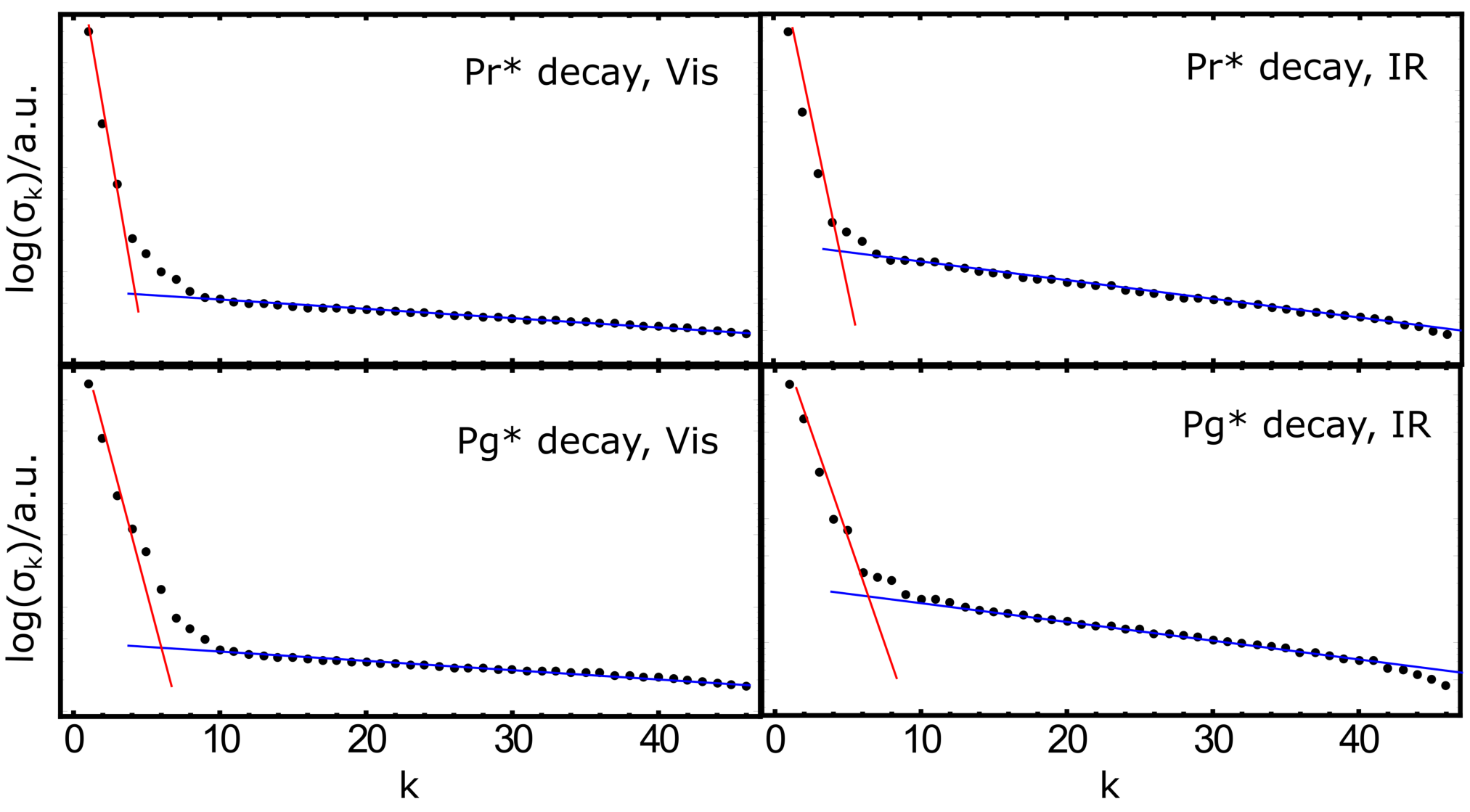}
	\caption{Singular values $\sigma_k$ of all data sets obtained from SVD plotted on a logarithmic scale. Lines guide the eye towards the noise level (points that fall on the blue lines) and non-noise singular values (red lines) and indicate which $\sigma_k$ cannot be clearly assigned.}
	\label{SVD}
\end{figure}

\subsection*{Global multiexponential fitting}
Global multiexponential fitting, in contrast, tries to minimize the number of exponential terms in Eq.~\ref{LDA}, and instead treats the corresponding time-constants $\tau_k$ as free fitting parameters. As a first step, in order to determine the number $n$ of exponentials needed to properly describe the data, all data sets were subjected to a singular value decomposition (SVD), as implemented in Wolfram Mathematica 12\cite{Mathematica,pre92}.  A SVD is a mathematical operation that uniquely decomposes the data matrix $d(\omega_i,t_j)$ in essence according to Eq.~\ref{separability}, it treats the $k$-columns of $C_k(t_j)$ and $A_k(\omega_i)$ as (unphysical) orthonormal vectors. It therefore introduces additional scaling factors $\sigma_k$ (the singular values):
\begin{equation}
d(\omega_i,t_j) = \sum_{k=1}^{n}C_k(t_j)\sigma_k A_k(\omega_i),
\label{SVD2}
\end{equation}
which describe the extent to which a particular component contributes to the overall signal. 

The resulting singular values are shown in Fig. \ref{SVD} on a logarithmic scale. Due to noise, SVD only provides an estimate for the minimal $n$, which is 3 for the decay of Pr* and 5 for Pg* for both spectral regions, i.e. the number of points that clearly fall on the red regression line in Fig. \ref{SVD}. There might be further components present, i.e., points that neither fall on the blue nor the red lines in figure \ref{SVD}, but they are close to the noise level of the data.

For the subsequent global multiexponential fitting, the matrix method for global exponential analysis was applied.\cite{Beckwith2018} Starting values for the time constants $\tau_k$ were determined from the peaks in the dynamical content (equation \ref{Dynamical content}). All data analysis was performed with home-written scripts in Wolfram Mathematica 12\cite{Mathematica}.

Lifetime analysis and multi-exponential fitting are complementary approaches with individual strengths and weaknesses. While global fitting describes the data with the smallest set of exponential functions necessary to account for a given complexity, lifetime analysis uses a quasi-continuum of exponential functions to detect maxima and minima in dynamics. While the global fitting is limited by number of exponentials, the regularisation in the lifetime analysis suppresses overfitting of rapidly changing small amplitudes.  Lifetime analysis offers the advantage that one does not have to pre-define the number of exponentials, and that it may also detect processes with small amplitudes. Global fitting, on the other hand, is directly related to the solutions of kinetic models that describe barrier-crossing phenomena between microscopic states with microscopic rate constants\cite{Beckwith2018,VanStokkum2004}.

\section*{Results}

\subsection*{Pr* decay, Vis spectral range }

\begin{figure}
	\centering
	\includegraphics[width=.5\linewidth]{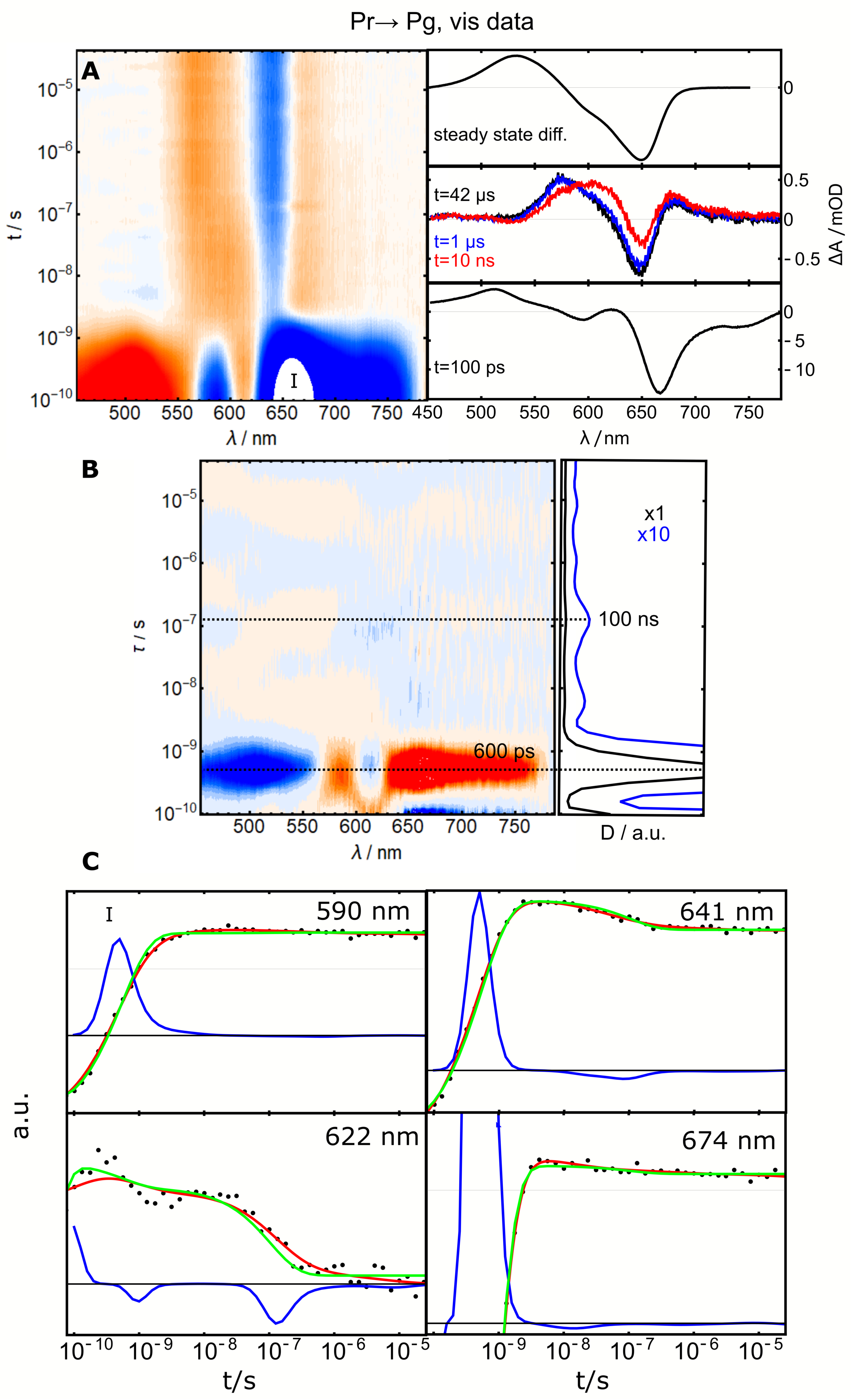}
	\caption{Transient Vis absorption data from 100~ps to 42~\textmu s after excitation of a Pr sample with 380~nm laser pulses. \textbf{A}: Color map representation of the data, red color code indicates positive absorbance change, blue negative change. Right panel: steady state Pg-minus-Pr difference spectrum and selected spectral traces at 100~ps, 10~ns, 1~\textmu s and 42~\textmu s. \textbf{B}: Lifetime spectrum and dynamical content $D(\tau_k)$ from lifetime analysis. The color code is defined such that the decay of a positive (red) signal in the raw data is represented here by blue amplitude and vice versa. Right panel: the dynamical content was scaled by a factor of x10 to reveal smaller features (blue line). Dashed horizontal lines are intended to guide the eye towards events local maxima in $D(\tau_k)$. \textbf{C}: Selected kinetic traces (black dots) with lifetime analysis fits (red lines), global fits (green lines) and amplitudes $a_{i,k}$ from lifetime analysis (blue lines).}
	
	\label{Fig2}
\end{figure}

Samples in the Pr state were excited with 380~nm laser pulses in the Soret band, and large transient signals (ca. 20~mOD) in the Vis were found at early times, which are associated with the  electronically excited state (Pr*, Fig. \ref{Fig2}A). This time window has been studied previously with high time resolution in the Vis after excitation in the Q-band between 620 and 670 nm, and identical results were obtained\cite{Slavov2015}. Different spectral regions were assigned to excited state absorption (ESA, mainly $<$550~nm), stimulated emission and ground state bleach (SE and GSB, strong negative band around 620-750~nm)\cite{Slavov2015}. The region in between (550-620~nm) contains overlapping contributions from ESA and GSB. Similar excited-state spectra were also obtained for several related red/green CBCRs from \textit{Nostoc punctiforme}, albeit with large variations in lifetimes \cite{Kim2012a,Gottlieb2015, Kirpich2019, Jenkins2019}. In the lifetime analysis of Slr-g3, a local maximum of $D$ is observed at 600~ps, and global fitting resulted in an optimized time constant of 540~ps ($\tau_1$, see table \ref{table1}). These values are in good agreement with the literature, where both lifetime analysis and global fitting produced similar results\cite{Slavov2015}. The lifetime spectra also display the previously reported smaller amplitudes with inverted sign before the dominant Pr* decay. This excited-state process is faster than 100~ps and may include contributions from the  excitation process, since it is at the edge of the time resolution of the present experiment (60~ps). In the global fits, a short time constant accounts for this signal ($\tau_0$). The overall similarity of the raw data, lifetime spectra and time constants for the Pr* decay reported here compared to the Q-band excitation experiments\cite{Slavov2015} provides solid evidence that the Pr* excited state dynamics are largely independent of the excitation wavelength.

After approximately 1~ns, Pr* vanishes and the remaining signals, which are roughly a factor of 10 smaller, can be safely assigned to ground state dynamics. Up to approximately 1~\textmu s, the ground state dynamics can be characterized mainly by a successive blue shift of the photoproduct absorption (Fig \ref{Fig2}A, right panels). We find no notable difference between the spectral traces at 1~\textmu s and the last transient at 42~\textmu s, indicating a meta-stable configuration that prevails during this time window. Both dynamical content and global fit yield a time constant of 100 ns for this process ($\tau_2$). Kinetic traces at selected wavelengths (Fig. \ref{Fig2}C) show that both the lifetime analysis and global fit represent the data reasonably well, albeit differing in details at later delay times, an effect that will be discussed below. Comparison of the last transient at 42~\textmu s to the steady-state Vis Pg-minus-Pr difference spectrum indicates that Pg is not yet formed within the time frame of the experiment (Fig. \ref{Fig2}A). This result is in good agreement with flash photolysis experiments of Slr-g3, which reported a similar spectrum with a broad maximum around 570~nm in the early microsecond range, while the formation of Pg proceeds with a much longer time constant of 1.1~ms \cite{Xu2014}.

\subsection*{Pr* decay, IR spectral range}

\begin{figure}
	\centering
	\includegraphics[width=.5\linewidth]{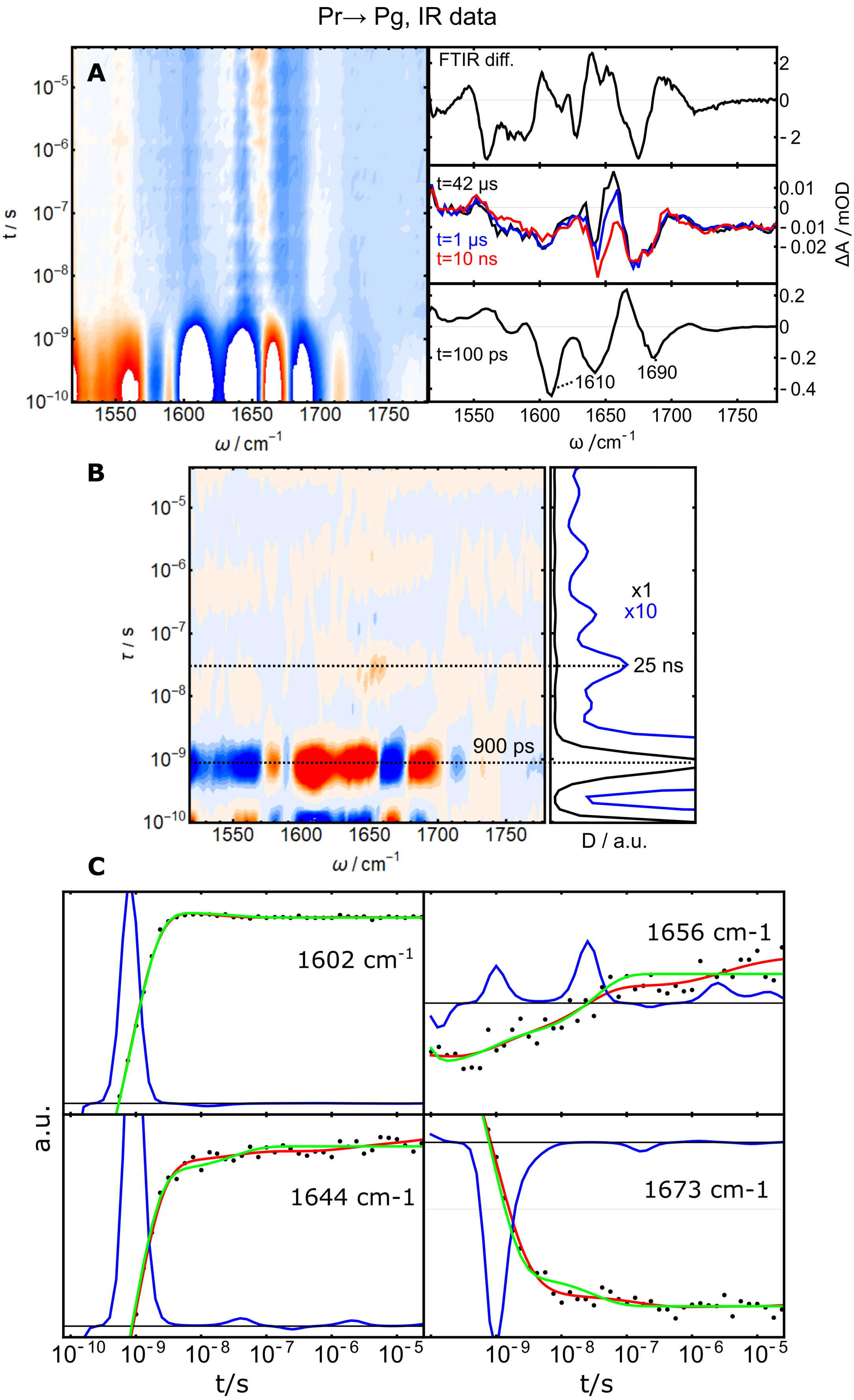}
	\caption{Transient IR absorption data from 100~ps to 42~\textmu s after excitation of a Pr sample with 380~nm laser pulses. \textbf{A}: Color map representation of the data. Right panel: steady state Pg-minus-Pr FT-IR difference spectrum and selected spectral traces at 100~ps, 10~ns, 1~\textmu s and 42~\textmu s. \textbf{B}: Lifetime spectrum and dynamical content $D(\tau_k)$. Right panel: the dynamical content (black and blue lines). Dashed horizontal lines are intended to guide the eye towards events local maxima in $D(\tau_k)$. \textbf{C}: Selected kinetic traces (black dots) with lifetime analysis fits (red lines), global fits (green lines) and amplitudes $a_{i,k}$ from lifetime analysis (blue lines).}
	\label{Fig3}
\end{figure}

IR difference spectra of cyanobacterial bili-proteins in the region between 1520 and 1780~cm$^{-1}$ are typically dominated by contributions from the PCB chromophore (C=C, C=N and C=O stretching modes) superimposed with amide I' signals originating from structural changes in the protein backbone \cite{VanThor2007, Fukushima2011}. Furthermore, C=O stretching vibrations from carboxylic groups can appear in this region. Due to the strong overlap of these different contributions, precise assignment of signals to specific vibrational modes is generally difficult, but some are possible based on literature. The steady-state IR difference spectra (Fig. \ref{intro}E) closely resemble the results obtained for the highly homologous red/green CBCR AnPixJ-g2 from \textit{Anabaena} PCC 7120 \cite{Fukushima2011, Song2015}. This related GAF domain was investigated by FT-IR spectroscopy with a uniformly $\mathrm{^{13} C,^{15} N}$-labeled PCB chromophore and thus some bands were assigned to PCB or the protein environment\cite{Song2015}. For example, a difference signal at 1690/1700~cm$^{-1}$  was assigned to the PCB D-ring carbonyl stretch. This finding is consistent with the present results: GSB signals at the respective frequencies are found in the picosecond data (Fig. \ref{Fig3} A), originating from the electronic excitation of the chromophore.
The strongest difference signal in the sub-ns traces lies around 1610~cm$^{-1}$, where a dominant negative band is commonly observed in other PCB-binding photoreceptors such as the cyanobacterial phytochrome Cph1 \cite{VanThor2007,Choudry2018} or the CBCRs Tlr0924 \cite{Hauck2014, Hardman2014} and TePixJ \cite{Hardman2020}. This ubiquitous feature can be assigned to the bleach of PCB C=C stretching signals upon electronic excitation. This C=C stretching mode overlaps only weakly with the amide I' region and can serve as a marker for the conjugated system of PCB in the transient IR data.

Similar to the Vis data, the IR spectra after excitation of Pr (Fig. \ref{Fig3}A) are dominated by excited state signals that decay with a time constant of 900~ps (dynamical content) or 860~ps (global fit, $\tau_1$). The slower kinetics of this process in the IR compared to the Vis are due to a kinetic isotope effect (KIE). The Vis spectra were also recorded in D$_2$O buffer (full Vis D$_2$O data set provided in the SI) and here the global fit yields a time constant of 1~ns (table \ref{table1}). A KIE in the excited state dynamics was previously reported for other bili-proteins and rationalized in terms of an excited state proton transfer \cite{Toh2010,Hontani2016}. Like in the Vis data, a ground state process in the nanosecond time regime is observed, mainly in the region between 1650 and 1670~cm$^{-1}$, but here the kinetics stretch into the microsecond range. This process is characterized by maximum dynamical content at around 25~ns (Fig. \ref{Fig3}B), while the global fit yields a time constant of 32~ns under the assumption that this time window is sufficiently described by one exponential ($\tau_2$). All prominent maxima of $D$ and corresponding time constants from the global fits for all Pg* decay data are summarized in table \ref{table1}.

\begin{table}
	\caption{Maxima of $D$ and time constants from the global fits with three components for all Pr* decay data. The short lived component $\tau_0$* is beyond the time resolution of the setup.}
	\begin{center}
	\begin{tabular}{|c|c|c|c|}
		\hline
		& $\tau_0$* / s& $\tau_1$ / s & $\tau_2$ / s \\
		\hline
		Vis, lifetime analysis $D$ & $<10^{-10}$& $6*10^{-10}$ & $1*10^{-7}$ \\
		\hline
		Vis, global fit&$2*10^{-11}$ & $5.4*10^{-10}$ &  $1*10^{-7}$\\
		\hline
		Vis, global fit D$_{2}$O &$1.6*10^{-11}$ & $1.0*10^{-9}$ &  $1.7*10^{-7}$\\
		\hline
		IR, lifetime analysis $D$ & $<10^{-10}$ &$9*10^{-10}$ & $2.5*10^{-8}$ \\
		\hline
		IR, global fit & $2.8*10^{-11}$ & $8.6*10^{-10}$  & $3.2*10^{-8}$ \\
		\hline
	\end{tabular}
	\label{table1}
\end{center}
\end{table}

\subsection*{Pg* decay, Vis spectral range}

\begin{figure}
	\centering
	\includegraphics[width=.5\linewidth]{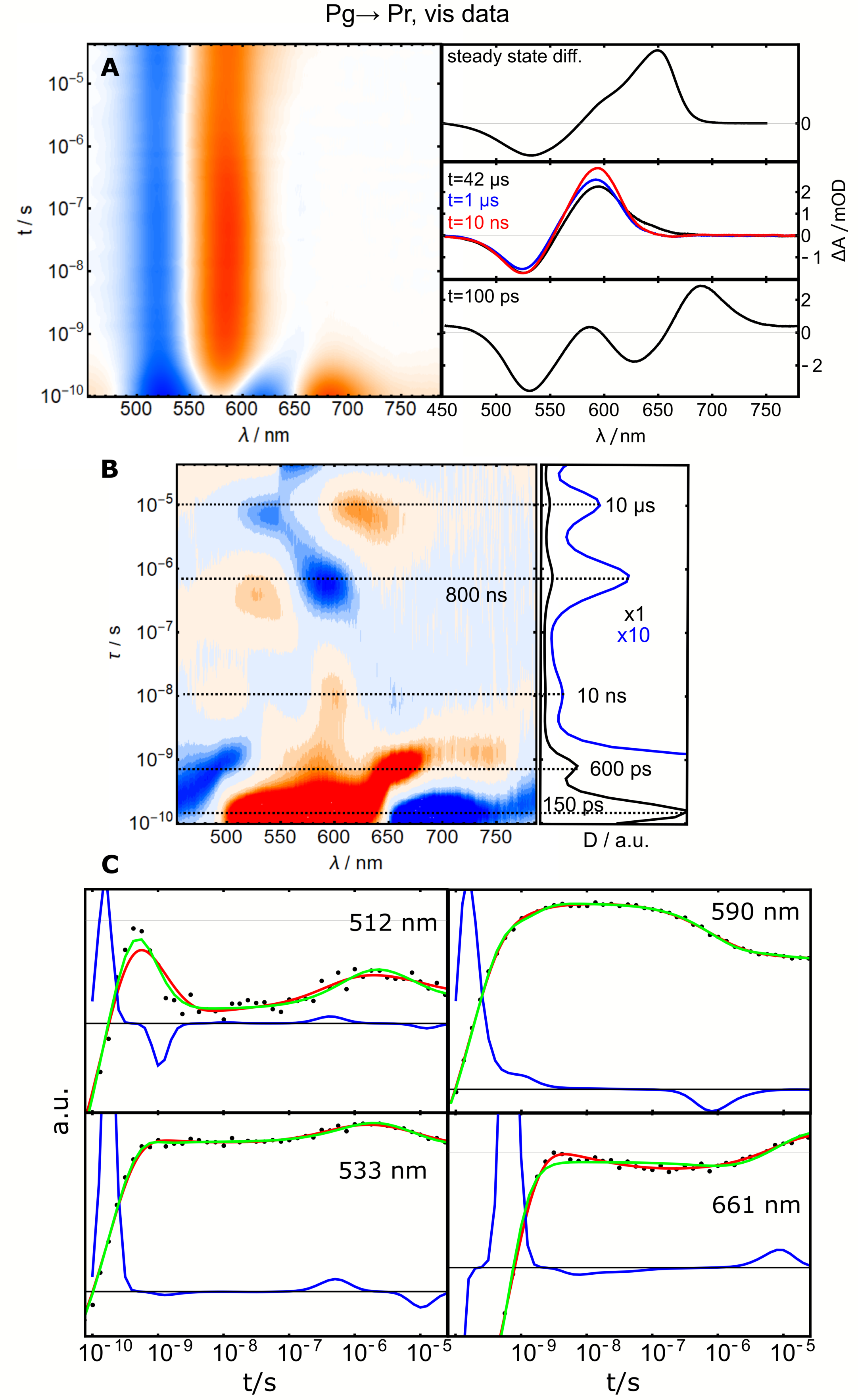}
	\caption{Transient Vis absorption data from 100~ps to 42~\textmu s after excitation of a Pg sample with 380~nm laser pulses. \textbf{A}: Color map representation of the data. Right panel: steady state Pr-minus-Pg difference spectrum and selected spectral traces at 100~ps, 10~ns, 1~\textmu s and 42~\textmu s. \textbf{B}: Lifetime spectrum and dynamical content $D(\tau_k)$. Right panel: the dynamical content (black and blue lines). Dashed horizontal lines are intended to guide the eye towards events local maxima in $D(\tau_k)$. \textbf{C}: Selected kinetic traces (black dots) with lifetime analysis fits (red lines), global fits (green lines) and amplitudes $a_{i,k}$ from lifetime analysis (blue lines).}
	\label{Fig4}
\end{figure}

The earliest signals after excitation of Pg samples (Fig. \ref{Fig4} A, 100~ps) are also in good agreement with the literature on Slr-g3 \cite{Slavov2015} and related red/green CBCRs \cite{Kim2012b, Kirpich2019a}. Here, the positive signal around 700~nm was assigned to ESA, while the negative features at 525 and 625~nm correspond to GSB and SE, respectively. By calculating the intensity ratio at 650~nm, we estimate ca. 10~\% residual Pr contribution in the Pg samples due to incomplete photoconversion by the background illumination procedure (Fig. \ref{intro}D). When these samples are excited at 380~nm (Soret band), the residual Pr contribution is also excited. However, compared to Pr*, Pg* decays faster and more productively, as judged from the relative intensities of the excited- and ground state related spectra (compare different scales in Fig. \ref{Fig2} and \ref{Fig4} A). The literature values for photochemical quantum yields are 0.08 (Pr* $\rightarrow$ Pg) and 0.3 (Pg* $\rightarrow$ Pr), respectively \cite{Pennacchietti2015}. Accordingly, two maxima are found at 150 and 600~ps in the dynamical content plots (Fig \ref{Fig4} B), which correspond to the decays of Pg* and Pr*, respectively. Note that this separation of the Pr*-related signals is not only true for the local maximum of $D(\tau_k)$, but also for the positive and negative amplitudes related to Pr* in the lifetime density map (compare to Figs \ref{Fig2} and \ref{Fig3} B). Thus, lifetime analysis allows a clear separation of the Pr* decay from Pg*.

In the subsequent ground state dynamics, a local maximum of the photoproduct absorption is found at 600~nm in the nanosecond range, before it decays and red-shifts at later times (Fig. \ref{Fig4} A, right panel). The dynamics on this time scale are evidence for further intermediate states that were not previously reported. Pr formation is not finished within the investigated time window and proceeds within 1~ms from an orange-absorbing intermediate \cite{Xu2014}. In total, five clearly separated maxima of $D$ are observed over the investigated time window (Fig. \ref{Fig4}B). Two of these kinetic processes correspond to the decay of Pg*  and Pr*, with time constants of $\tau_1$=140~ps and $\tau_2$=605~ps, respectively (150 and 600~ps from dynamical content). The literature reports a time constant of 111~ps for the decay of Pg* after excitation in the Q-Band, which is in reasonable agreement with the present data\cite{Slavov2015}. Three long time constants $\tau_3$-$\tau_5$ of 11~ns, 710~ns and 8.6~\textmu s are obtained from the global fit (after optimization starting from 10~ns, 800~ns and  10~\textmu s), and are associated with ground state dynamics that follow the decay of Pg* (maxima of $D$). The ground state dynamics that are expected from the parallel decay of the 10\% residual Pr* are too small to contribute on this scale (we estimate a factor of 50).  Figure \ref{Fig4} C shows that both fitting methods work well and do not deviate significantly from the data.

\subsection*{Pg* decay, IR spectral range}

\begin{figure}
	\centering
	\includegraphics[width=.5\linewidth]{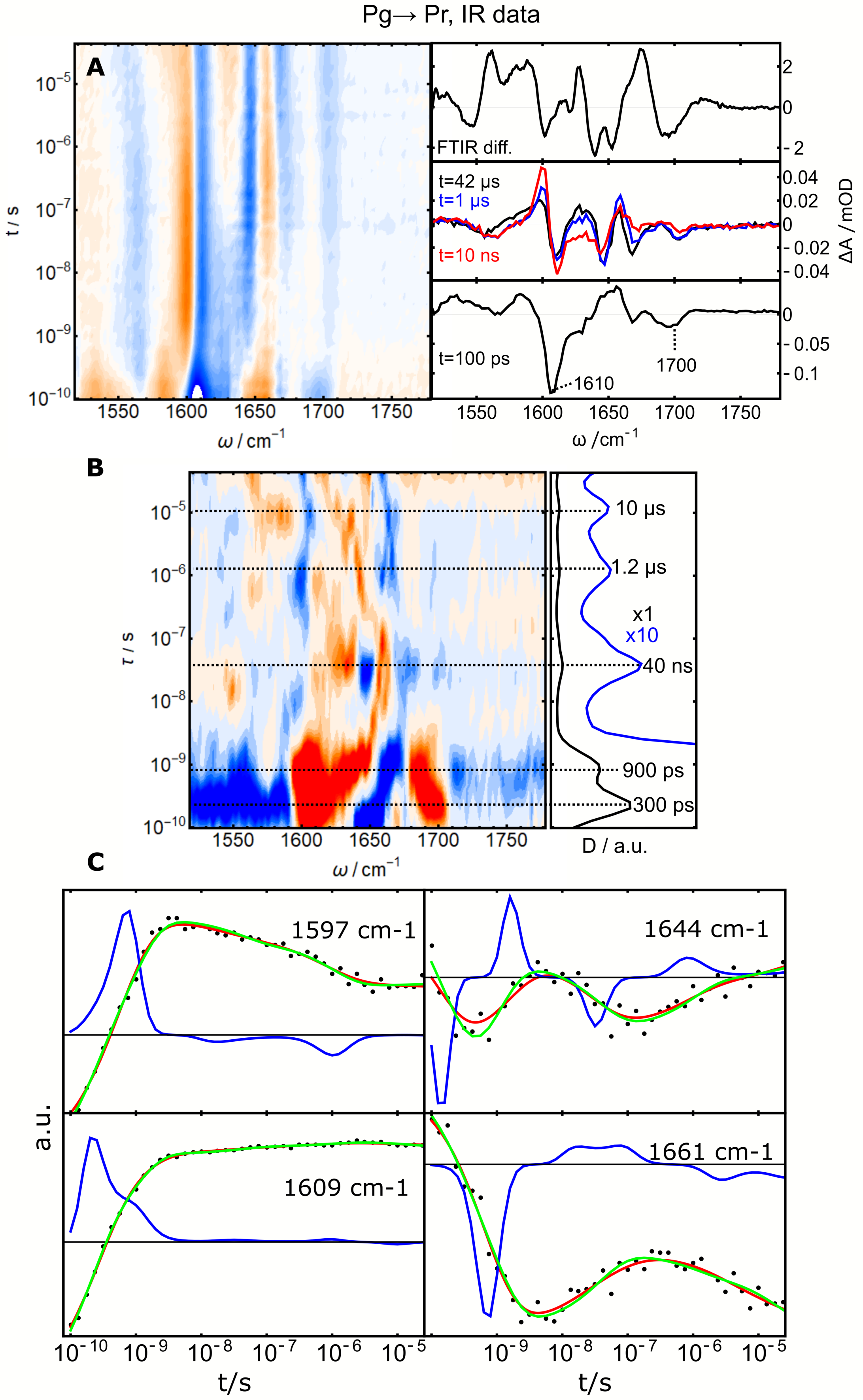}
	\caption{Transient IR absorption data from 100~ps to 42~\textmu s after excitation of a Pg sample with 380~nm laser pulses. \textbf{A}: Color map representation of the data. Right panel: steady state Pr-minus-Pg FT-IR difference spectrum and selected spectral traces at 100~ps, 10~ns, 1~\textmu s and 42~\textmu s. \textbf{B}: Lifetime spectrum and dynamical content $D(\tau_k)$. Right panel: the dynamical content (black and blue lines). Dashed horizontal lines are intended to guide the eye towards events local maxima in $D(\tau_k)$. \textbf{C}: Selected kinetic traces (black dots) with lifetime analysis fits (red lines), global fits (green lines) and amplitudes $a_{i,k}$ from lifetime analysis (blue lines).}
	\label{Fig5}
\end{figure}

Like in the Vis region, the transient IR spectra display kinetics associated with at least five non-noise singular values (points that fall on the red line in Fig.\ref{SVD}) and five corresponding local maxima of dynamical content (Fig. \ref{Fig5}B). Like in the Vis data, two well-separated dominant time constants in the sub-nanosecond regime describe the decay of the excited states. Pg* decays with $\tau_1$=300~ps in lifetime analysis (180~ps from global fit) and Pr* with 900 ps (930~ps from global fit), which is in good agreement with the Pr* excitation experiment ($\tau_1$ in table \ref{table1}).
Later, $\tau_3$ is detected around 40~ns in the lifetime spectra (Fig. \ref{Fig5}B). Here, strong amplitudes appear in the amide I' region between 1620 and 1680~cm$^{-1}$, and a small feature is observed at 1600~cm$^{-1}$ slightly earlier around 10~ns, while the global fit yields an optimized global time constant of $\tau_3$=38~ns in this time range. Two additional time constants $\tau_4$ and $\tau_5$ appear at 1.2 and 10 \textmu s in the lifetime spectra and are optimized to 1 and 20~\textmu s in the global fit. These two time constants match very well with $\tau_4$ and $\tau_5$ from the Vis data, indicating that the same processes are monitored in both spectral windows. All time constants that are associated with the decay of Pg* are summarized in table \ref{table2}.

\begin{table*}[tbhp]
	\caption{Maxima of $D$ and time constants from the global fits with five components for all Pg* decay data. $\tau_2$* is associated with the parallel decay of Pr* due to the excitation scheme.}
\begin{center}
	\begin{tabular}{|c|c|c|c|c|c|}
		\hline
		& $\tau_1$ / s & $\tau_2$* / s & $\tau_3$ / s & $\tau_4$ / s & $\tau_5$ / s\\
		\hline
		Vis, lifetime analysis $D$& $1.5*10^{-10}$ &  $6*10^{-10}$ & $1*10^{-8}$ & $8*10^{-7}$ & $1*10^{-5}$\\
		\hline
		Vis, global fit& $1.4*10^{-10}$ & $6*10^{-10}$ & $1.1*10^{-8}$ & $7.1*10^{-7}$ & $8.6*10^{-6}$\\
		\hline
		Vis, global fit D$_2$O& $1.1*10^{-10}$ & $8.9*10^{-10}$ & $1.8*10^{-8}$ & $7.3*10^{-7}$ & $1.3*10^{-5}$\\
		\hline
		IR,lifetime analysis $D$& $3*10^{-10}$ & $9*10^{-10}$ & $4*10^{-8}$ & $1.2*10^{-6}$ & $1*10^{-5}$\\
		\hline
		IR, global fit&  $1.8*10^{-10}$ &  $9.3*10^{-10}$ & $3.8*10^{-8}$& $1*10^{-6}$ & $2*10^{-5}$\\
		\hline
	\end{tabular}
	\label{table2}
\end{center}
\end{table*}

\section*{Discussion}

\subsection*{Ground state Intermediates}

The decay of of the electronically excited states Pr* and Pg* are by far the dominant processes and are thus associated with large difference signals, but subsequent ground-state processes are also resolved. Singular value decomposition and both fitting methods indicate that there is at least one nanosecond ground-state process subsequent to Pr* decay and at least three processes following the decay of Pg* in the investigated time window. The macroscopic time constants obtained from global fitting generally agree with the positions of local maxima of the dynamical content in the lifetime analysis within less than a factor of two.

In the classical kinetic picture, $n$ observed processes are associated with $n+1$ states, with unique, time-independent spectra and microscopic rate constants in a kinetic matrix. This picture is similar to a Markov-State model, where the time constants are related to the elements of the transition matrix. With the time constants extracted from global fitting or the lifetime analysis as only direct source of information, it is unfortunately not possible to construct a kinetic model or directly derive the microscopic rate constants that constitute the kinetic matrix, because the problem is inherently underdetermined. To see why that is the case, one has to consider the $n \times n$ kinetic matrix of a system with $n$ states. The fact that the rates out of each state need to enter some other states introduces constraints, which reduce the number of independent matrix elements to $n \times (n-1)$. That matrix only has $n-1$ non-zero eigenvalues, which are the observed timescales ($\tau_k$). However, that information is not sufficient to determine all $n \times (n-1)$ elements of the kinetic matrix. Applied to the three time scales found after the Pg* decay ($\tau_3 - \tau_5$, table \ref{table2}, $n-1=3$), we would need to determine 12 independent entries of a 4x4 kinetic matrix only in this time window, which is impossible without further assumptions.

In this situation, it is common practice to \textit{a priori} assume a specific photocycle model, in which most of the entries of the kinetic matrix are set to zero and the other entries are ordered in a unique way. Evolution-associated difference spectra (EADS) builds on such a model, constructing a unidirectional sequential reaction pathway with the time constants ordered from fast to slow\cite{VanStokkum2004}. Closely related to such sequential models is the common nomenclature for the ground state intermediates of bilin-photoreceptors. It was originally derived from cryo-trap Vis absorption experiments on \textit{Avena sativa} phytochrome A and denotes the primary ground-state photoproducts as "Lumi" and the subsequent thermal intermediates as "Meta" states \cite{Eilfeld1985}. This naming scheme was extended several times for different systems, e.g. early/late Lumi \cite{Kuebel2020} or several cases of Meta sub-states \cite{Information2015, Schmidt2018, Buhrke2020a} that are distinguishable with different spectroscopic methods. Following this naming scheme, we would need to introduce two more metastable ground-state intermediates after the decay of Pr* (Lumi-R and Meta-R1), and four after Pg* (Lumi-G and Meta-G1 to 3), based on the number of time-constants we identified in the investigated time window $<$42 \textmu s. However, several recent studies show that ground state dynamics in bilin photoreceptors are generally non-unidirectional and non-sequential\cite{Takala2018, Buhrke2018a}, and especially in CBCRs, photochemical heterogeneity was reported for several systems including Slr-g3 \cite{Lim2018,Slavov2015, Buhrke2020}. Therefore we think that the decomposition of the presented spectroscopic data into EADS would be misleading.

\subsection*{Protein adjustments in the nanosecond regime}

It was recently shown that the Vis spectra of Slr-g3 are selective and sensitive probes of the effective conjugation length of PCB\cite{Wiebeler2018, Wiebeler2019, Xu2020}, while IR spectra are expected to contain contributions from both PCB and protein environment. Thus, all events that show up strongly in the Vis data likely include adjustments of dihedral angles between the pyrrole rings, which have a large impact on the effective conjugation length and also affect IR signals of the chromophore, e.g. the C=C stretching in the region around 1610~cm$^{-1}$. For such processes, we expect to find similar time constants in the Vis and IR spectra, which is indeed true for most of the data. However, if features associated with certain time scales appear dominant in IR transients in the amide I' region with only very minor counterparts in the Vis, they likely originate from structural changes in the protein surroundings and have only minor impact on the conjugated system of PCB. Such a pattern is found around 40~ns after the decay of Pg* (table \ref{table2},~$\tau_3$), indicating adjustments of the protein surroundings on that time scale that are only weakly sensed by the chromophore and thus have only minor contributions in the Vis. This is an interesting finding because the time scale from 1-1000 ns is only sparsely studied by IR spectroscopy, mainly due to experimental challenges. Here, step-scan FT-IR experiments are not very sensitive and quite time consuming, while ultrafast methods are limited by the length of optical delay lines to a few nanoseconds. An alternative approach is the external synchronisation of separate pump and probe laser sources like in this study and a few others. One recent IR study of \textit{Deinococcus Radiodurans} phytochrome in this time window also identified a complex kinetic behaviour related to metastable intermediate states \cite{Kuebel2020}, while no significant protein kinetics in the nanosecond range were found in an algal phytochrome\cite{Choudry2018} and a blue/green CBCR\cite{Hardman2020}. In the present case of a red/green CBCR, we find a number of intermediate states in the nanosecond regime that involve structural rearrangements of the protein surroundings. These rearrangements are potentially involved in signal transduction to a catalytic domain and the allosteric regulation process, which appears to be very variable in different bilin-photoreceptors\cite{Jenkins2019,Gourinchas2019a}.

\subsection*{Comparison of fitting methods}

\begin{figure}
	\centering
	\includegraphics[width=.5\linewidth]{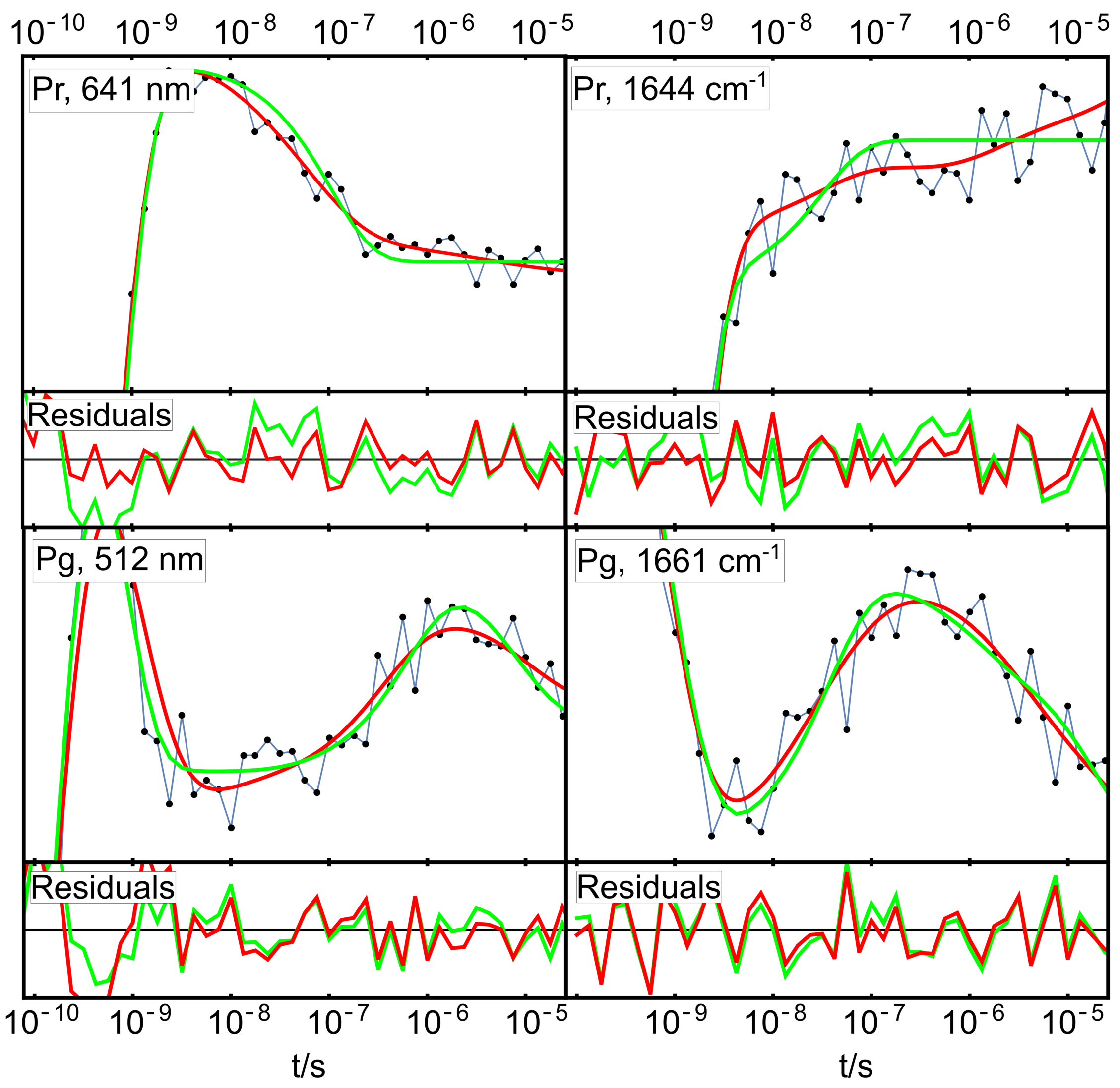}
	\caption{Detailed view of selected kinetic traces (black dots) with lifetime analysis fits (red lines) and global fits (green lines). The regular residuals of the two fits are plotted below the graphs.}
	\label{zoom}
\end{figure}

A fundamental assumption behind multiexponential global fitting is that the system under investigation can be described by classical kinetics, i.e. interconverting states with discrete, time-invariant spectra (equation \ref{separability})\cite{VanStokkum2004}. The solution of such a system is a sum of exponentials. The assumption of classical kinetics is well justified for the decay of electronic excited states that involve large energy differences between well separated electronic states and large scale local structural and spectral changes. On longer time scales, the kinetics are related to thermal processes with considerably smaller energy barriers. An established model in protein folding studies is the diffusion of the system on a rugged energy landscape\cite{Onuchic1997,Dill2008,Osvath2003}. The thermal fluctuations of thousands of atoms contribute to the non-equilibrium evolution of the amide I' signal in the IR spectra and can also influence the Vis absorption of the chromophore. Therefore, one could expect that thermal fluctuations should add a distribution-character to the multi-exponential kinetics. Such an effect could be captured by the lifetime analysis, but not the global fit. Figure \ref{zoom} shows detailed views of kinetic traces from the different data sets (more in the SI). We generally observe that the global multiexponential fits and lifetime analysis produce inflection points at the same positions. One difference is that the slope of the global fits is generally steeper than the exponential distribution from the lifetime analysis. However, the differences are small and lie within the error of the noisy data (see residuals). We therefore conclude that lifetime analysis and global fits are equivalent alternative representations of noisy data sets.   

\section*{Conclusion}

Altogether, we find rich kinetics in the Vis and IR regions upon excitation of Slr-g3 samples in the Pr and Pg states with 380~nm laser pulses. The earliest processes that we observe are related to the decay of the electronically excited states Pr* and Pg* and are in good agreement with the literature\cite{Slavov2015}. At later times up to 42 \textmu s we observe further ground-state processes that can be related to at least two metastable intermediate states that follow the decay of Pr* and four intermediates formed from Pg*. We extract the macroscopic time-constants $\tau_k$ of these processes with two fitting methods and find good agreement between them.

\subsection{Supplementary Material}

The supplementary material to this manuscript includes steady state Vis spectra before and after the lyophilisation procedure, absolute FTIR spectra, a section about regularisation with the l-curve criterion including l-curves, regularisation parameters and alternative lifetime maps, Vis TA data measured in D$_2$O buffer including the same fitting and representation as all other data in the manuscript, a figure that reveals the KIE in the raw data and further plots of kinetic traces with the different fits and residuals. 

\subsection{Author contributions}

DB and PJH expressed and purified the protein. DB, KTO and RFT implemented and optimised set-ups and performed spectroscopic experiments. DB and PH analyzed the data. DB and PH wrote the manuscript with contributions from all authors.

\subsection{Acknowledgements}

The authors thank Thomas Friedrich and co-workers for providing the Slr-g3 expressing \textit{E.coli} cell line, Olga Božović for help with protein expression, Roland Zehnder and Jan Helbing for technical support and Friedrich Siebert and Chavdar Slavov for helpful comments. This work was supported by the Swiss National Science Foundation (SNF) through the NCCR MUST and Grant 200020B\_188694/1.

\subsection{Data availability statement}

The data that support the findings of this study are available from the corresponding author upon reasonable request.

\bibliographystyle{aip}
\bibliography{refs}

\end{document}